\documentclass[letterpaper, 10 pt, conference]{ieeeconf}  

\IEEEoverridecommandlockouts                              

\usepackage{graphics} 
\usepackage{epsfig} 
\usepackage{mathptmx} 
\usepackage{times} 
\usepackage{amsmath} 
\usepackage{amssymb}  
\usepackage{url}
\usepackage{graphicx}
\usepackage{subcaption} 
\usepackage[most]{tcolorbox}
\newtcolorbox[auto counter,number within=section]{promptbox}[2][]{%
  colback=gray!3,colframe=gray!60!black,boxrule=0.6pt,arc=2pt,
  fonttitle=\bfseries,title={Box~\thetcbcounter: #2},#1}

\title{\LARGE \bf
A Review of Large Language Models for Stock Price Forecasting from a Hedge-Fund Perspective*\thanks{*Accepted at the IEEE Conference on Artificial Intelligence, Spain, May 8--10, 2026.}
}

\author{Olivia Zhang$^{1}$ and Zhilin Zhang$^{2}$
\thanks{$^{1}$Olivia Zhang is with the Hockaday School, 11600 Welch Rd, Dallas, TX 75229, USA
        (e-mail: ozhang30@hockaday.org)}%
\thanks{$^{2}$Zhilin Zhang is with Lumos Alpha,
        Dallas, TX 75248, USA
        (e-mail: zhilinzhang@lumosalpha.com)}%
}

\begin{document}

\maketitle
\thispagestyle{empty}
\pagestyle{empty}

\begin{abstract}
Large language models (LLMs) are increasingly deployed in quantitative finance for stock price forecasting. This review synthesizes recent applications of LLMs in this domain, including extracting sentiment from financial news and social media, analyzing financial reports and earnings-call transcripts, tokenizing or symbolizing stock price series, and constructing multi-agent trading systems. Particular attention is paid to practical pitfalls that are often understated in the literature, such as fragility in sentiment analysis, dataset and horizon design, performance evaluation metrics, data leakage, illiquidity premia, and limits of stock price predictability. Organized from a hedge-fund perspective, the review is intended to guide both academic researchers and hedge fund managers in integrating LLMs into real-world trading pipelines and in stress-testing their robustness under realistic market frictions.
\end{abstract}


\section{Introduction}

Machine learning–based stock price forecasting has attracted sustained interest across finance and machine learning communities \cite{DePrado2018AFML,EnhancingLiteratureReview}. Among recent advances, large language models (LLMs) have emerged as a versatile class of methods with promising results in this application. While several surveys review LLMs in finance broadly \cite{SurveyofLLMs,LLMsInFinance,LLMsForFinancialInvestment,ReviewDatasetsCaseStudy,LLMinEquityMarkets}, comprehensive coverage devoted specifically to LLMs for stock price forecasting remains limited. Moreover, existing reviews are largely academic-oriented, with comparatively little emphasis on the perspectives and practical considerations from hedge funds.   

This review fills that gap by synthesizing recent advances in LLM-driven stock price forecasting and assessing open issues, challenges, and practical considerations through a hedge-fund lens. It surveys major areas in LLMs for stock price forecasting: (1) sentiment extraction from news and social media, (2) factual analysis of financial reports and earnings-call transcripts, (3) tokenization and symbolization of price series, and (4) multi-agent trading systems. Then it examines  pitfalls in the literature, including shortcomings in sentiment modeling, dataset construction, performance evaluation, data-leakage controls, and illiquidity handling, followed by a discussion of stock price predictability. Many of these considerations are underemphasized in prior surveys and often overlooked in academic algorithm design, yet they are critical for robust, deployable trading pipelines in real markets.

The contributions of this review are threefold. First, it proposes a focused review of LLM roles in stock price forecasting and trading systems. Second, it provides a critical assessment of empirical practices in the current literature from the perspective of hedge fund managers, highlighting where academic benchmarks might diverge from realistic deployment conditions. Third, it distills practical guidelines and open research questions for integrating LLMs into production-grade trading pipelines and for stress-testing their robustness under market frictions, thereby supporting both academic researchers and practitioners working at the intersection of AI and quantitative finance.

\section{Usage of LLMs in Stock Price Forecasting}

LLMs have been applied to stock price forecasting in several roles: (i) sentiment extraction; (ii) analysis of financial reports and earnings-call transcripts; (iii) discovery of relationships across stocks, markets, and assets; (iv) processing of tokenization/symbolization of price series; and (v) design of agentic AI trading architectures. The following subsections review each area in turn.

\subsection{Sentiment Analysis}
The goal of sentiment analysis on financial markets is to infer investors' attitudes and evaluations toward a company or a financial market. It has been an active research topic for years \cite{CriticalReview,FinancialSentimentAnalysis,Du2024LLMFinSent}. Recently, with the emerging and innovation of LLMs, researchers have been actively developing techniques to use LLMs to extract sentiment-related information from news articles, financial reports, or social media posts \cite{ReviewDatasetsCaseStudy,GPTInvestAR,Ploutos,WallStreetNeophyte,CanChatGPTforecast,TemporalMeetsLLM,UsingFinancialNews,LinkingMicroblog,TransformingSentiment,Risklabs,TemporalCNN,WhatDoLLMsKnow,ForecastingCryptocurrencyReturns,EnhancingAutomatedTrading}. The sentiment-related information can be used as features for downstream machine learning models in stock price prediction or portfolio construction, or used through zero- or few-shot prompting for stock movement prediction. Moreover, it can be used as supervision to fine-tune pretrained LLMs for task-specific market forecasts. This paradigm leverages LLMs’ strengths in language understanding and contextual reasoning, enriching the information set beyond prices alone and enabling more accurate market inferences.

A common use of LLMs for sentiment analysis is to classify news or news headlines of stocks into multiple emotion categories (e.g. positive, negative, or neutral) or sentiment scores, which are further converted to trading signals \cite{CanChatGPTforecast,Ploutos,TemporalMeetsLLM,UsingFinancialNews,LinkingMicroblog,GoodDebtBadDebt,TemporalCNN,WhatDoLLMsKnow,DeepLearningAndNLP}. For example, in \cite{CanChatGPTforecast} a model classifies the positive or negative sentiment of news headlines, and the results are directly transformed to long or short decisions on stocks. The work found that the predictive power is strongest for negative news and small-cap stocks, and is most pronounced within the first two days after the news release. In \cite{Ploutos} the authors fine-tune a LLaMA-2 model in the finance domain to classify company-specific news over a five-day window as positive/neutral/negative and aggregate these labels into a sentiment signal. In \cite{TemporalMeetsLLM} the authors prompt a GPT-4 model to summarize each article and extract keywords, then compress the generated data into a weekly “meta” summary that captures the tone and significant drivers around each stock. The summaries, alongside prices and company profiles, function as the sentiment signal that is fed to the model. In \cite{UsingFinancialNews}, researchers present a finetuned FinBERT \cite{FinBERTAraci2019} model that calculates sentiment scores on news articles. The scores become features to a random forest model that predicts stock price direction for the next day. 


When using LLMs to extract or classify sentiment from news or social media posts, careful prompt engineering is important especially in zero-shot modes or few-shot modes \cite{Du2024LLMFinSent}. In \cite{LinkingMicroblog} the authors use careful prompt design to extract sentiment from blogs, resulting in improved daily stock direction prediction. The work in \cite{TransformingSentiment} further shows that with the right prompts, ChatGPT in zero-shot modes outperforms FinBERT on headline sentiment and produces signals that closely align with the market's next-day returns. 

Besides prompt engineering, there are also studies on how to label sentiment. The work in \cite{CBITS} points out that news sentiment should be labeled based on how the event is likely to affect price. Therefore, if an article discussing a stock is unlikely to impact its price, its label should be neutral even if the overall tone is positive or negative. In other words, sentiment labeling is price movement-focused. In \cite{EnhancingFewShot} the authors propose an alternative way to improve the predictability of sentiment by filtering irrelevant news across relevant headlines with a few-shot LLM framework.


After using LLMs to extract sentiment information, the sentiment can be further processed as features \cite{LLMFeatureExtraction,LLMGuided,CrossSectorMarketRegime} for machine learning models, or used directly for downstream market forecasting and decision-making. For example, Mudarisov et al. \cite{CrossSectorMarketRegime} propose an LLM-augmented framework, combining data from news and time-series. It scores “market-mover” articles, maps sentiment with a FinBERT model to next-period regime probabilities, and fuses these with a scaler to forecast market regimes, significantly outperforming time-series-only baselines.

Most recently, researchers have proposed multi-agent LLMs \cite{DesigningHeterogeneousLLM,FinSentLLM} for the sentiment classification task. LLM-based multi-agent models can better extract sentiment and generate sentiment signals, especially related to market trends and deeper causal relationships. The topic of agentic trading systems will be discussed more thoroughly later.

\subsection{Financial Report and Earnings-Call Analysis}

Additionally, LLMs can extract predictive features from financial reports, analyst reports, and earnings-call transcripts \cite{GPTInvestAR,ECCAnalyzer,FromAnalystReports,CombiningFinancialDataNews}. However, instead of inferring the sentiment in texts, this method aims to identify relevant facts. In \cite{GPTInvestAR}, researchers use a GPT-3 model to analyze 10-K filings of companies, saving huge amounts of time that professionals and investment firms would otherwise have needed. The proposed GPT-InvestAR model answers 27 fixed prompts relating to each company report and gives confidence scores from 0 to 100, which are used as features for a Linear Regression model to rank stock price movement.

A multimodal pipeline that encodes earnings-call audio and transcript texts is proposed in \cite{ECCAnalyzer}. It uses hierarchical LLM summarization and a “question bank” to extract predictive features for short-term volatility prediction. The work in \cite{CombiningFinancialDataNews} constructs a retrieval-augmented few-shot LLM framework, which jointly ingests financial statements, price-momentum features, and summarized news to classify three to six months stock directions.


\subsection{Extraction of Relationships of Stocks, Markets, and Assets}
In addition to sentiment and predictive features, one can use LLMs to reveal other contextual information, such as relationships among stocks, market sectors, or assets, for improved price forecasting performance. In \cite{ChatGPTInformedGraph}, Chen et al. builds a ChatGPT model to read each day's financial headlines. They identify which Dow-30 stocks are co-affected and therefore the relationships between companies, producing a time-varying co-affect graph. Trained on the graph, a neural network combined with an LSTM and MLP classifies next-day stocks' movement directions. As a result, the method yields higher cumulative returns and lower drawdowns than Deep Learning-based benchmarks. The work in \cite{LLMKnowledgeEnhancement} uses an LLM to automatically extract relation triplets among market indicators, fuses this knowledge with numeric signals into a heterogeneous graph, and predicts market index trends with a knowledge-enhanced Heterogeneous Graph Transformer, outperforming baselines.

\subsection{Tokenization/Symbolization of Stock Prices}

Utilizing the pattern recognition capabilities in LLMs, time series numeric values can be first discretized into intervals or bins \cite{SAX} and then mapped to tokens (or embeddings) \cite{StockGPT,LLMZeroshot,TemporalMeetsLLM,ChatGPTInformedGraph,TimeLLM}. The resulting token sequence is used as input to an LLM. In this setup, time series forecasting is represented as simply next-token prediction of the token sequence. Because LLMs excel at capturing long-range dependencies and repetitiveness in text, this technique enables models to learn seasonality, trend, and autocorrelation patterns, which may be highly relevant in predicting future stock prices.

Besides capturing important patterns in time series, this approach may have many other advantages. It leverages models to learn price patterns more directly instead of through news articles that may occasionally contain biases or misinformation \cite{BuyTesla}. Also, it ensures that predictions are less dependent on the quantity of accessible news data, allowing for on-the-spot predictions and trading. Finally, unlike sentiment, researchers can use quantization to predict multiple-step future returns, rather than a single point or direction \cite{StockGPT}.

However, the performance of LLMs using tokenization or symbolization of time series is sensitive to the binning design and pre-processing procedures. The number of bins, normalization, differencing, and spacing between tokens can change what a model "sees" and different models may react differently to the same modification. For example, \cite{LLMZeroshot} reveals that while inserting spaces between digits to alter token boundaries improves performance in zero-shot GPT-3 models, LLaMA-2 models are built to tokenize digits by default, so performance using the same method on LLaMA-2 models degrades. Therefore, bin edges, scaling, and tokenization should be model-specific hyperparameters and are best not transferred across different model architectures.

Many studies on the tokenization or symbolization of stock data have been published. \cite{StockGPT} discretizes daily stock return data into bins numbered from 0 to 401. The processed return data is then used to train a GPT model, which makes daily forecasts of returns from each stock. It goes long on the top decile of stocks with the highest return forecast and short on the bottom decile. Under equal weighting, the portfolio yields an average annual return of 119\% and a Sharpe ratio of 6.5. However, the portfolio's ultra-high performance is subject to contraints, as detailed later in this paper. \cite{TemporalMeetsLLM} also converts numerical stock price data of each Nasdaq-100 company into symbol sequences such as D4, D5, U4, and U5. The symbolized data is then fed into the model, along with GPT4-generated news summaries and key words relating to the company. In \cite{TimeLLM} the authors normalize the input time series, convert it into embedding patches, and reprogram the patches into text prototype representations that are more natural for zero-shot or few-shot LLMs. They provide guidance for the model's reasoning with a Prompt-as-Prefix prompt, which gives input statistics and clearer instructions on the task. The paper yields an average of over 23.5\% and 12.4\% reduction in the mean square error (MSE) and the mean absolute error (MAE) across  baseline models. Moreover, \cite{LLMZeroshot} proposes methods for effective tokenization of time series values as a string of numerical digits and trains zero-shot GPT-3 and LLaMA-2 models on the tokenized data. The authors show that zero-shot large language models with the correct tokenization processes can even exceed the performance of models trained specifically for downstream tasks.

Altogether, LLMs using this approach have been shown to be successful, as they are able to directly model the temporal dynamics of stock price data and capture essential seasonality and trend patterns. While this approach has achieved excellent results, it is highly sensitive to preprocessing design and model-specific hyperparameters. There is still considerable improvement that can be made to this area in the future.

\subsection{Multi-agentic Decision Systems}

With the recent advances in Agentic AI \cite{FromLLMtoAIAgents,AdvancingInnovation}, a surge of researchers are beginning to develop such autonomous systems for stock market trading \cite{TradingR1}. There are two types of agentic systems: single agent and multi-agent. For single-agent models, one agent, though perhaps with multiple tools and modules, handles the inputs, reasoning, and decision-making and iteratively learns and refines itself. 
Multi-agent LLMs will instantiate "roles" (e.g. fundamental, sentiment, and technical analysts, risk evaluators, etc.) that assess a given text from different perspectives and produce structured judgments \cite{DesigningHeterogeneousLLM}. These outputs are then aggregated into portfolio actions or financial market prediction. This review will focus more on multi-agentic decision systems.

One advantage of multi-agent trading systems \cite{TradingR1,TradingAgents,MultimodalFinAgent} is that they are well-rounded. They can fuse the knowledge from each specialist into a single, well thought-out decision. This can  reduce single-model hallucinations and overfitting, as the agents must argue with each other and weigh their respective merits before arriving at a consensus. Many agentic systems \cite{MultimodalFinAgent} are also able to combine multi-modal data, such as text, stock prices, and stock charts. In addition, multi-agent systems permit each agent to be optimized in parallel for a narrower task, which usually outperforms having one monolithic model and makes training more efficient.

On the other hand, AI agents are very reliant on data to make accurate predictions and so may be sensitive to the quality of information they receive \cite{AdvancingInnovation}. In the real world, though, financial data can be incredibly noisy, nonstationary, and inconsistent. Therefore, the prediction of some models can be unreliable. Moreover, the transparency and ethicality of AI agents are also crucial aspects for the practical application of agentic trading systems. Agents need to be able to comply with regulatory standards and explain their actions if necessary.

A growing body of recent work applies multi-agent LLM architectures to stock price forecasting. For instance, \cite{FinArena} and \cite{EnhancingInvestment} adopt a classic structure of multi-agent systems, Mixture-of-Expert (MoE), where specialized modules analyze different data types and an "universal" module aggregates the signals together. In the work \cite{FinArena}, which introduces a model called FinArena, there is a trio of agents: Time Series Agent, News Agent, and Statement Agent, and a final agent that takes into account investors' risk preferences and makes the decision. In \cite{EnhancingInvestment}, a similar structure is proposed; a multi-agent collaboration system analyzes SEC 10K forms of 30 companies on three aspects--- fundamentals, market sentiment, and risk analysis. \cite{AlphaAgents} presents a system where a Valuation Agent, Sentiment Agent, and Fundamental Agent debate on stock recommendations. And, \cite{IntegratingTraditionalTechnical} uses the concept of Elliot Waves as well as seven agents that communicate in a chain to identify patterns in market conditions. 
In \cite{FromNewsToForecast} a multi-agent, reflection-driven LLM framework is proposed. It retrieves and filters news, tags short- vs long-term impacts, and uses an evaluation agent to iteratively refine the selection logic. The selected events and contextual data are then prompt-integrated and used to fine-tune an LLM that forecasts the next time-series tokens. Collectively, these studies highlight the potential of multi-agentic decision systems in revolutionizing investment analysis.

Agentic trading systems exhibit several characteristics and typically operate in a perceive–reason–act cycle. During the perception phase, the systems first interpret the task specification (e.g., prompts or objectives) and ingest relevant data sources, preparing structured inputs for subsequent reasoning and decision-making.
To reduce hallucinations, paradigms such as Retrieval-Augmented Generation (RAG) are recommended \cite{RetrievalAugmentedGeneration}. RAG is a method where a language model answers a query by first fetching related, real-time facts from external sources such as the internet and databases, and generates a response citing these facts. This technique can help the systems tie answers to actual facts. It also allows the systems to pull the latest information without fine-tuning. During the perception stage, memory is needed, too. Short-term memory can preserve immediate context while long-term memory stores huge amounts of past experience needed to improve the systems.

A specialized structure is also essential to encourage the agents to devise complex reasoning, consider multiple scenarios, and perform self-critique. Popular paradigms that can be plugged into agentic trading loops include ReACT for interleaving reasoning with tool use \cite{ReAct}, Chain-of-Thought (CoT) for stepwise deduction \cite{Chain-of-thought}, Tree-of-Thought (ToT) for exploring alternate reasoning \cite{Tree-of-thoughts}, and Reflexion for self-critique and iterative improvement \cite{Reflexion}.

Finally, engineering frameworks such as LangChain or LangGraph \cite{LangGraph}, AutoGen \cite{AutoGen}, and CrewAI \cite{CrewAI} are useful for building agentic trading systems. They provide tools and libraries for developers and hedge funds to deploy agentic AI.

\section{Issues and Challenges from a Hedge-Fund Perspective}

The following section examines the issues and challenges of applying LLMs to stock price forecasting under real-world trading constraints. Although prior studies \cite{ReviewDatasetsCaseStudy,CriticalReview,ComparativeAnalysis,LLMinEquityMarkets,IsAllInfo,Du2024LLMFinSent,WhatDoesChatGPT} have noted related concerns, they are largely framed from an academic perspective. In contrast, this review critically evaluates the practical feasibility of LLM-based trading systems and identifies the specific pitfalls that emerge during implementation.

\subsection{Issues in Sentiment Analysis}

There is a well-known behavioral finance  experiment conducted by a wealth management firm~\cite{ELMgame,WSJ} which granted participants an almost god-like ``Crystal Ball'' preview of the next day’s \emph{Wall Street Journal} front page. Participants could take directional bets—going long the S\&P~500 when they believed headlines would be bullish and shorting when bearish. Yet even with this near-omniscient edge, results were weak: the hit rate was 51.5\%; approximately 45\% of participants lost money, and 16\% went bust—underscoring that foreknowledge of news does not, by itself, translate into consistent profits.

Some machine learning–based trading approaches show similarly modest results. For instance, \cite{WallStreetNeophyte} documents hit rates only marginally above chance and observes that LLMs often surface superficial correlations rather than causal structure. Similarly, the use of sentiment as a predictive signal for stock returns has been criticized for limited robustness and stability \cite{CriticalReview,IsAllInfo}.

A key reason sentiment-only signals struggle to predict market direction is their strong context dependence: the \emph{same} news can imply opposite market directions under different macro regimes, policy backdrops, or prior expectations. For example, a ``stronger-than-expected payrolls'' report may be bullish in a disinflationary, soft-landing regime, yet bearish in an overheating regime if it increases the perceived likelihood of tighter monetary policy and higher interest rates. 

Prominent market participants can further destabilize the effectiveness of news-based sentiment signals. For example, trading news related to Cathie Wood, the founder and portfolio manager of the ARK Innovation Fund, appeared positively correlated with the performance of growth stocks prior to 2022\footnote{During 2020–2021, financial TV channels and forums were saturated with coverage of Cathie Wood and her ARK ETFs.}, but exhibited a less or negative association during 2022–2023. Such sign reversals highlight nonstationarity and reflexivity in sentiment–return relationships. Consequently, a sentiment-driven model trained predominantly on pre-2022 news is likely to generalize poorly thereafter, underscoring the need for explicit regime detection, periodic re-calibration, and robustness checks at deployment.

In most studies, sentiment is extracted from news and user-generated content such as blogs, Twitter/X, and Reddit posts \cite{LinkingMicroblog,TemporalCNN,WhatDoLLMsKnow,ForecastingCryptocurrencyReturns,DeepLearningAndNLP}. However, little work evaluates the data quality of different textual sources. Content from professional investment outlets, such as Seeking Alpha (\url{https://seekingalpha.com/}) and Zacks (\url{https://www.zacks.com/}), may produce more predictive sentiment signals than general-purpose social media, due to domain expertise and editorial standards. This hypothesis, however, requires rigorous evaluation before solid conclusions can be drawn.

In summary, without explicit conditioning on the macroeconomic regime, prior expectations, market positioning, and policy reaction functions, sentiment signals can be unstable over extended horizons. This fragility is compounded by heterogeneity across textual sources and by reflexive dynamics \cite{SorosReflexivity} tied to prominent market participants. Yet, most LLM-based sentiment pipelines neither model these contextual variables nor control for source quality and market participant-specific effects, yielding regime- and source-dependent mappings from text to returns and, consequently, fragile out-of-sample performance. Moreover, potential biases arising from source selection, survivorship and publication effects, platform-specific moderation, and the NLP pipeline itself (e.g., sampling, deduplication, tokenization, labeling, and prompt framing) remain underexplored and require systematic study.




\subsection{Issues in Datasets}

Warren Buffett commands exceptional respect in the investment community, owing largely to his sustained, multi-decade track record of compounding returns at roughly 20\% annually. By contrast, it is not difficult to find individuals or strategies that outperform over one- or two-year windows, but the short-term outperformance rarely earns the same respect. The distinction underscores a central tenet of asset management: only strategies that deliver durable, repeatable performance across market cycles, rather than transient and period-specific gains, merit credibility as genuine investment approaches.

Hence, a widely accepted practice in hedge funds is to test a strategy over at least one full market cycle including both bull and bear phases. Yet most LLM-based studies conduct evaluation over very short horizons, failing to cover a complete cycle. Consequently, reported gains may reflect regime-specific effects rather than robust, out-of-sample generalization.

For example, widely used datasets such as BigData22~\cite{BigData22} and CIKM18~\cite{CIKM18} each span roughly one calendar year, whereas ACL18~\cite{ACL18} covers only about two years. Algorithms trained on these corpora are often evaluated over even shorter horizons. Moreover, studies relying on self-constructed datasets sometimes report results based on only a few weeks or months of data. Such limited windows can reduce statistical power, heighten sample-period and regime-dependence biases, and undermine the credibility of reported performance for real-world deployment.

Admittedly, assembling a comprehensive, long-horizon dataset that integrates financial text corpora with historical market data is challenging and will require sustained collaboration between the investment and academic communities.


\subsection{Issues in Performance Evaluation}

In many studies, newly proposed LLM algorithms are evaluated only against LLM baselines. A more informative assessment would include comparison to established, non-LLM trading methods (e.g., machine learning or technical analysis strategies~\cite{TAbook}). Indeed, the performance gains reported for some LLM algorithms appear attainable with relatively simple rules based on standard technical analysis (TA) indicators.

Some works do incorporate the rules based on TA indicators into the evaluation, but the associated parameter choices are often ad hoc or random. Furthermore, the predictive content of most TA indicators resides primarily in their dynamics (e.g., slopes, crossovers, and sign changes) rather than in their  values. Rigorous evaluations should therefore employ well justified parameterizations, conduct robustness checks across plausible grids, and emphasize directional or state-transition signals (e.g., first differences, threshold crossings) in addition to raw indicator values.

Some other work compares their LLM-based methods solely to a buy-and-hold benchmark. However, as noted in the previous subsection, outperforming buy-and-hold over a short horizon is not particularly informative.

With respect to evaluation metrics, MSE, MAE, F$_1$ score, and accuracy are common in general time series forecasting but are poorly aligned with real-world trading objectives. For example, if today’s closing price of a stock is used as the predicted value for tomorrow’s closing price, the resulting MSE will often be very small, yet such a ``prediction'' is meaningless for an actual trading system. In addition, these metrics reward pointwise closeness of price levels rather than meaningful trading signals, and can therefore overstate the practical value of a model in live deployment.

From a hedge-fund perspective, risk-adjusted performance is more important and practical. Therefore, it is preferable that evaluations report profit and loss, Sharpe and Sortino ratios, maximum drawdown, turnover, and capacity, among others. While some recent studies have begun to adopt several of these metrics, their use remains far from universal.

\subsection{Data Leakage}

Using sentiment extracted from news and social media posts to forecast next-day or next-week stock returns is highly susceptible to future information leakage, especially when a single event generates posts over multiple days. Some corporate news unfold over even several months, such as acquisitions and mergers, product launches, and market expansions. Some common leakage situations include:

\begin{itemize}
  \item \textbf{Cross-day event drift.} A firm's earnings announcement may be discussed for 2--3 days. Posts made a few days after the earnings announcement will embed market reaction details. So, if day-$(t{+}1)$ or day-$(t{+}2)$ posts are included as features for predicting day $t$  via random train/test splits, the outcome leaks into the model.
  
  
  \item \textbf{Literal label statements.} When constructing datasets, texts that explicitly state realized price movements can leak the target label if included before labeling. Examples include: ``TSLA tumbled after the announcement,'' ``AAPL surged 5\% into the close,'' or ``META rebounded at midday when the news broke.'' In such cases, the model is not inferring from predictive signals; it is merely reading the answer. To avoid this leakage, one needs to exclude retrospective summaries or enforce time filters that remove texts revealing contemporaneous or realized price changes prior to the labeling horizon.

  \item \textbf{Peer and supply-chain spillover.} Major events can diffuse to suppliers and competitors with lags. For instance, sentiment around Intel’s earnings has historically provided clues about Advanced Micro Devices’s short-horizon movements days later; including such lagged posts can tacitly encode future information about the peer.
\end{itemize}

Mitigating leakage requires careful, event-centric preprocessing and time-respecting evaluation protocols, such as constructing event-level features, assigning all posts from the same event to a single fold, and using forward-only (rolling/expanding) splits with strict timestamp cutoffs. 

Admittedly, such data preparation is extremely complicated. But without these safeguards, reported performance is likely to be biased.

\subsection{Illiquidity and Premiums}

Many published studies, particularly those involving daily long–short strategies in small- and mid-cap stocks, implicitly assume sufficient liquidity. In practice, though, these stocks often have relatively low trading volumes. For hedge funds and institutional managers, executing at the modeled price is often infeasible: orders incur slippage and market impact, creating a persistent transaction premium over the notional entry/exit. Even small daily shortfalls compound materially. 

Consider a simple illustration over 252 trading days (approximately one calendar year)\footnote{For simplicity, this example subtracts from transaction costs and other frictions beyond the stated premium.}. Under sufficient liquidity, a strategy that compounds at 0.30\% per day attains a total return of
$(1+0.003)^{252}-1 \approx 112.73\%$.
Now introduce a 10\,bps daily execution premium due to illiquidity. This reduces the effective daily return to 0.20\%. The total return then becomes
$(1+0.002)^{252}-1 \approx 65.45\%$.
Hence, even a seemingly modest daily drag of 10\,bps due to illiquidity can materially erode cumulative performance over a one-year horizon. A similar erosion is documented in \cite{CanChatGPTforecast}: after accounting for a 0.2\% cost in daily trading, the proposed strategy becomes unprofitable.

Note that assuming a 0.1\% illiquidity premium is reasonable. For example, within the Russell 2000 (a market-cap-weighted index of roughly 2{,}000 U.S. small-cap equities), the mean and median of such premiums across component stocks can be around 0.2\% and 0.1\%, respectively.\footnote{Premiums were computed over the final 30 minutes before the market close on each trading day from October 6 to October 10, 2025, and then averaged across days.} When a hedge fund deploys substantial capital to build a position, the illiquidity premium can become even more pronounced. Therefore, ignoring these frictions overstates implementable returns and strategy capacity.


\subsection{Predictability of Stock Prices}

There is a long-standing debate over whether stock price data follow a random walk. Some studies support the random-walk hypothesis \cite{RandomWalk}, while others present evidence to the contrary \cite{NonRandomWalk}. A pragmatic view is that markets alternate between regimes: in some periods prices behave like a drifted random walk, and in others they deviate from it. 

This suggests that an approach centered on regime-switch detection \cite{CrossSectorMarketRegime} followed by machine learning–based trading may be effective (i.e., first forecasting non-random-walk intervals and applying trading models, while holding or abstaining from trading if forecasting random-walk intervals). Most recent LLM-based approaches to stock price forecasting do not incorporate explicit regime handling. Multi-agent LLM frameworks may be better suited to this problem, with specialized agents for regime identification, feature extraction, forecasting, and risk management. However, regardless of which technique is best suited for stock price prediction, one should be cautious against naïvely using LLMs \cite{WhatDoesChatGPT,IsAllInfo}.


\section{Conclusion}

This review surveyed recent use of large language models for stock price forecasting, evaluated existing methods and their strengths and limitations, and highlighted practical considerations from a hedge-fund perspective. Although often underemphasized in academic studies, these issues are central to real-world trading. The analysis underscores the gap between academic prototypes and real-world trading systems and points out several directions for future work.

In particular, from a real-world implementation perspective, the survey suggests the following practical guidelines for LLM-based stock price forecasting:
\begin{itemize}
  \item \textbf{Use long-horizon, regime-spanning datasets.} Evaluate models over at least one full market cycle that includes both bull and bear phases, rather than short, favorable windows.

  \item \textbf{Align metrics with trading objectives.} Place greater emphasis on profits and losses and risk-adjusted performance (Sharpe, Sortino, drawdown, turnover, capacity), while using statistical error metrics such as MSE or accuracy as secondary indicators.

  \item \textbf{Control data leakage rigorously.} Employ event-level or forward-only splits, time-aligned features, and deduplication to prevent future information from contaminating training data.

  \item \textbf{Account for illiquidity, costs, and capacity.} Incorporate realistic frictions, slippage, and position-size limits into backtests, especially for small- and mid-cap stocks.

  \item \textbf{Benchmark against transparent non-LLM baselines.} Include well-specified, technical analysis-based trading systems and non-LLM machine learning models with clearly documented configurations and robustness checks.
\end{itemize}










\begin{thebibliography}{99}

\bibitem{DePrado2018AFML}
M.~L.~de~Prado, \textit{Advances in Financial Machine Learning}. John Wiley \& Sons, 2018.


\bibitem{EnhancingLiteratureReview} S. {\L}aniewski and R. {'S}lepaczuk, ``Enhancing literature review with LLM and NLP methods: Algorithmic trading case,'' \textit{arXiv preprint} arXiv:2411.05013, 2024.

\bibitem{SurveyofLLMs} Y. Nie, Y. Kong, X. Dong, \emph{et al.}, ``A survey of large language models for financial applications: Progress, prospects and challenges,'' \textit{arXiv preprint} arXiv:2406.11903, 2024.

\bibitem{LLMsInFinance} Y.~Li, S.~Wang, H.~Ding, and H.~Chen, ``Large language models in finance: A survey,'' in \textit{Proceedings of the Fourth ACM International Conference on AI in Finance}, pp. 374--382, 2023.

\bibitem{LLMsForFinancialInvestment} Y. Kong, Y. Nie, X. Dong, \emph{et al.}, ``Large language models for financial and investment management: Applications and benchmarks,'' \textit{Journal of Portfolio Management}, vol. 51, no. 2, 2024.

\bibitem{LLMinEquityMarkets} A. Jadhav and V. Mirza, ``Large language models in equity markets: Applications, techniques, and insights,'' \textit{Frontiers in Artificial Intelligence}, vol. 8, Art. 1608365, 2025. doi: 10.3389/frai.2025.1608365.

\bibitem{CriticalReview} Z. Jankov{'a}, ``Critical review of text mining and sentiment analysis for stock market prediction,'' \textit{Journal of Business Economics and Management}, vol. 24, no. 1, pp. 177--198, 2023.

\bibitem{FinancialSentimentAnalysis} K. Du, F. Xing, R. Mao, and E. Cambria, ``Financial sentiment analysis: Techniques and applications,'' \textit{ACM Computing Surveys}, vol. 56, no. 9, pp. 1--42, 2024.

\bibitem{GPTInvestAR} U. Gupta, ``GPT\hbox{-}InvestAR: Enhancing stock investment strategies through annual report analysis with large language models,'' \textit{arXiv preprint} arXiv:2309.03079, 2023.

\bibitem{StockGPT} D. Mai, ``StockGPT: A GenAI Model for Stock Prediction and Trading,'' \textit{The Journal of Financial Data Science}, early access, Sep. 17, 2025. doi: 10.3905/jfds.2025.1.202.

\bibitem{Ploutos} H. Tong, J. Li, N. Wu, \emph{et al.}, ``Ploutos: Towards explainable stock movement prediction with financial large language model,'' in \textit{Companion Proceedings of the ACM on Web Conference 2025}, pp. 490--499, 2025.

\bibitem{EnhancingAutomatedTrading} M. T. Siddique, S. S. Jamee, A. Sajal, \emph{et al.}, ``Enhancing automated trading with sentiment analysis: Leveraging large language models for stock market predictions,'' \textit{The American Journal of Engineering and Technology}, vol. 7, no. 3, pp. 185--195, 2025.


\bibitem{WallStreetNeophyte} Q. Xie, W. Han, Y. Lai, \emph{et al.}, ``The Wall Street Neophyte: A zero-shot analysis of ChatGPT over multimodal stock movement prediction challenges,'' \textit{arXiv preprint} arXiv:2304.05351, 2023.

\bibitem{CanChatGPTforecast} A. Lopez\hbox{-}Lira and Y. Tang, ``Can ChatGPT forecast stock price movements? Return predictability and large language models,'' \textit{arXiv preprint} arXiv:2304.07619, 2023.

\bibitem{TemporalMeetsLLM} X. Yu, Z. Chen, Y. Ling, \emph{et al.}, ``Temporal data meets LLM—Explainable financial time series forecasting,'' \textit{arXiv preprint} arXiv:2306.11025, 2023.

\bibitem{UsingFinancialNews} B. Fazlija and P. Harder, ``Using financial news sentiment for stock price direction prediction,'' \textit{Mathematics}, vol. 10, no. 13, pp. 2156, 2022.

\bibitem{LinkingMicroblog} R. Steinert and S. Altmann, ``Linking microblogging sentiments to stock price movement: An application of GPT\hbox{-}4,'' \textit{arXiv preprint} arXiv:2308.16771, 2023.

\bibitem{TransformingSentiment} G. Fatouros, J. Soldatos, K. Kouroumali, \emph{et al.}, ``Transforming sentiment analysis in the financial domain with ChatGPT,'' \textit{Machine Learning with Applications}, vol. 14, pp. 100508, 2023.

\bibitem{Risklabs} Y. Cao, Z. Chen, Q. Pei, \emph{et al.}, ``Risklabs: Predicting financial risk using large language model based on multi-sources data,'' \textit{Technical Report}, 2024.

\bibitem{FinBERTAraci2019} D. Araci, ``FinBERT: Financial sentiment analysis with pre-trained language models,'' \textit{Technical Report}, arXiv:1908.10063, 2019.

\bibitem{LLMKnowledgeEnhancement}
Z. Di, J. Chen, Y. Yang, \emph{et al.}, ``LLM-Driven Knowledge Enhancement for Securities Index Prediction,'' in \textit{Proceedings of LKM@IJCAI}, 2024.


\bibitem{TemporalCNN} C. M. Liapis and S. Kotsiantis, ``Temporal convolutional networks and BERT-based multi-label emotion analysis for financial forecasting,'' \textit{Information}, vol. 14, no. 11, pp. 596, 2023, MDPI.

\bibitem{FinSentLLM} Z. Zhang, R. Fu, Y. He, \emph{et al.}, ``FinSentLLM: Multi-LLM and structured semantic signals for enhanced financial sentiment forecasting,'' \textit{arXiv preprint} arXiv:2509.12638, 2025.

\bibitem{LLMFeatureExtraction} S. H. Saffarian and S. Haratizadeh, ``LLM-driven feature extraction for stock market prediction: A case study of Tehran Stock Exchange,'' in \textit{Proceedings of the 2024 15th International Conference on Information and Knowledge Technology (IKT)}, pp. 59--65, 2024.

\bibitem{LLMGuided} O. Mutian, J. J. Thomas, Y. Tianzhou, and U. Fiore, ``LLM-guided semantic feature selection for interpretable financial market forecasting in low-resource financial markets,'' \textit{Franklin Open}, p. 100359, 2025.



\bibitem{ECCAnalyzer} Y. Cao, Z. Chen, Q. Pei, \emph{et al.}, ``ECC Analyzer: Extracting trading signal from earnings conference calls using large language model for stock volatility prediction,'' in \textit{Proceedings of the 5th ACM International Conference on AI in Finance}, pp. 257--265, 2024.

\bibitem{FromAnalystReports} A. Moreno and J. Ordieres-Mer{'e}, ``Predicting stock price trends using language models to extract the sentiment from analyst reports: Evidence from IBEX 35-listed companies,'' \textit{Economics Letters}, p. 112404, 2025.


\bibitem{CombiningFinancialDataNews} A. Elahi and F. Taghvaei, ``Combining financial data and news articles for stock price movement prediction using large language models,'' in \textit{Proceedings of the 2024 IEEE International Conference on Big Data (BigData)}, pp. 4875--4883, 2024.

\bibitem{ComparativeAnalysis} M. Abdelsamie and H. Wang, ``Comparative analysis of LLM-based market prediction and human expertise with sentiment analysis and machine learning integration,'' in \textit{Proceedings of the 2024 7th International Conference on Data Science and Information Technology (DSIT)}, pp. 1--6, 2024.

\bibitem{CrossSectorMarketRegime} T. Mudarisov, R. V. State, Z. Kraussl, \emph{et al.}, ``Cross-sector market regime forecasting with LLM-augmented news analysis,'' in \textit{Proceedings of the 5th ACM International Conference on AI in Finance}, pp. 461--468, 2024.


\bibitem{ELMgame} V. Haghani and J. White, ``When a Crystal Ball Isn’t Enough to Make You Rich,'' \textit{Available at SSRN}, 2024.

\bibitem{WSJ} S. Jakab, ``Would a Time Machine Make You a Great Investor?,'' The Wall Street Journal, Oct. 14, 2024. [Online]. Available: https://www.wsj.com/finance/investing/would-a-time-machine-make-you-a-great-investor-7a4b39b8

\bibitem{CBITS} G. Kim, M. Kim, B. Kim, and H. Lim, ``CBITS: Crypto BERT incorporated trading system,'' \textit{IEEE Access}, vol. 11, pp. 6912--6921, 2023.

\bibitem{EnhancingFewShot} Y. Deng, X. He, J. Hu, and S.-M. Yiu, ``Enhancing few-shot stock trend prediction with large language models,'' \textit{arXiv preprint} arXiv:2407.09003, 2024.



\bibitem{WhatDoLLMsKnow} X. Deng, V. Bashlovkina, F. Han, \emph{et al.}, ``What do LLMs know about financial markets? A case study on Reddit market sentiment analysis,'' in \textit{Companion Proceedings of the ACM Web Conference 2023}, pp. 107--110, 2023.

\bibitem{ForecastingCryptocurrencyReturns} D. Ider and S. Lessmann, ``Forecasting cryptocurrency returns from sentiment signals: An analysis of BERT classifiers and weak supervision,'' \textit{arXiv preprint} arXiv:2204.05781, 2022.

\bibitem{DeepLearningAndNLP} V. Gurgul, S. Lessmann, and W. K. H{"a}rdle, ``Deep learning and NLP in cryptocurrency forecasting: Integrating financial, blockchain, and social media data,'' \textit{International Journal of Forecasting}, Elsevier, 2025.

\bibitem{GoodDebtBadDebt} P. Malo, A. Sinha, P. Korhonen, \emph{et al.}, ``Good debt or bad debt: Detecting semantic orientations in economic texts,'' \textit{Journal of the Association for Information Science and Technology}, vol. 65, no. 4, pp. 782--796, 2014.


\bibitem{LLMZeroshot} N. Gruver, M. Finzi, S. Qiu, and A. G. Wilson, ``Large Language Models Are Zero-Shot Time Series Forecasters,'' in \textit{Advances in Neural Information Processing Systems}, vol. 36. Red Hook, NY: Curran Associates, 2023, pp. 19622--19635.

\bibitem{ChatGPTInformedGraph} Z. Chen, L. N. Zheng, C. Lu, \emph{et al.}, ``ChatGPT-informed graph neural network for stock movement prediction,'' \textit{arXiv preprint} arXiv:2306.03763, 2023.

\bibitem{TimeLLM} M. Jin, S. Wang, L. Ma, \emph{et al.}, ``Time-LLM: Time series forecasting by reprogramming large language models,'' \textit{arXiv preprint} arXiv:2310.01728, 2023.


\bibitem{BuyTesla} C. Chuang and Y. Yang, ``Buy Tesla, sell Ford: Assessing implicit stock market preference in pre-trained language models,'' in \textit{Proceedings of the 60th Annual Meeting of the Association for Computational Linguistics (Volume 2: Short Papers)}, pp. 100--105, 2022.

\bibitem{ReviewDatasetsCaseStudy} C. Liu, A. Arulappan, R. Naha, \emph{et al.}, ``Large language models and sentiment analysis in financial markets: A review, datasets and case study,'' \textit{IEEE Access}, 2024.


\bibitem{SAX} J. Lin, E. Keogh, L. Wei, and S. Lonardi, ``Experiencing SAX: A novel symbolic representation of time series,'' \textit{Data Mining and Knowledge Discovery}, vol. 15, no. 2, pp. 107--144, 2007.

\bibitem{AdvancingInnovation} S. Joshi, ``Advancing innovation in financial stability: A comprehensive review of AI agent frameworks, challenges and applications,'' \textit{World Journal of Advanced Engineering Technology and Sciences}, vol. 14, no. 2, pp. 117--126, 2025.

\bibitem{FinArena} C. Xu, Z. Liu, and Z. Li, ``FinArena: A human-agent collaboration framework for financial market analysis and forecasting,'' \textit{arXiv preprint} arXiv:2503.02692, 2025.

\bibitem{EnhancingInvestment} X. Han, N. Wang, S. Che, \emph{et al.}, ``Enhancing investment analysis: Optimizing AI-agent collaboration in financial research,'' in \textit{Proceedings of the 5th ACM International Conference on AI in Finance}, pp. 538--546, 2024.

\bibitem{AlphaAgents} T. Zhao, J. Lyu, S. Jones, \emph{et al.}, ``AlphaAgents: Large language model based multi-agents for equity portfolio constructions,'' \textit{arXiv preprint} arXiv:2508.11152, 2025.

\bibitem{DesigningHeterogeneousLLM} F. Xing, ``Designing heterogeneous LLM agents for financial sentiment analysis,'' \textit{ACM Transactions on Management Information Systems}, vol. 16, no. 1, pp. 1--24, 2025, ACM, New York, NY.

\bibitem{IntegratingTraditionalTechnical} M. Wawer and J. A. Chudziak, ``Integrating traditional technical analysis with AI: A multi-agent LLM-based approach to stock market forecasting,'' \textit{arXiv preprint} arXiv:2506.16813, 2025.

\bibitem{RetrievalAugmentedGeneration} P. Lewis, E. Perez, A. Piktus, \emph{et al.}, ``Retrieval-augmented generation for knowledge-intensive NLP tasks,'' \textit{Advances in Neural Information Processing Systems}, vol. 33, pp. 9459--9474, 2020.

\bibitem{ReAct} S. Yao, J. Zhao, D. Yu, \emph{et al.}, ``ReAct: Synergizing reasoning and acting in language models,'' in \textit{Proceedings of the Eleventh International Conference on Learning Representations}, 2022.

\bibitem{Chain-of-thought} J. Wei, X. Wang, D. Schuurmans, \emph{et al.}, ``Chain-of-thought prompting elicits reasoning in large language models,'' \textit{Advances in Neural Information Processing Systems}, vol. 35, pp. 24824--24837, 2022.

\bibitem{Tree-of-thoughts} S. Yao, D. Yu, J. Zhao, \emph{et al.}, ``Tree of thoughts: Deliberate problem solving with large language models,'' \textit{Advances in Neural Information Processing Systems}, vol. 36, pp. 11809--11822, 2023.

\bibitem{Reflexion} N. Shinn, F. Cassano, A. Gopinath, \emph{et al.}, ``Reflexion: Language agents with verbal reinforcement learning,'' \textit{Advances in Neural Information Processing Systems}, vol. 36, pp. 8634--8652, 2023.

\bibitem{LangGraph} J. Wang and Z. Duan, ``Agent AI with LangGraph: A modular framework for enhancing machine translation using large language models,'' \textit{arXiv preprint} arXiv:2412.03801, 2024.

\bibitem{AutoGen} Q. Wu, G. Bansal, J. Zhang, \emph{et al.}, ``AutoGen: Enabling next-gen LLM applications via multi-agent conversations,'' in \textit{Proceedings of the First Conference on Language Modeling}, 2024.


\bibitem{CrewAI} Z. Duan and J. Wang, ``Exploration of LLM multi-agent application implementation based on LangGraph+ CrewAI,'' \textit{arXiv preprint} arXiv:2411.18241, 2024.



\bibitem{TradingR1} Y. Xiao, E. Sun, T. Chen, \emph{et al.}, ``Trading-R1: Financial trading with LLM reasoning via reinforcement learning,'' \textit{arXiv preprint} arXiv:2509.11420, 2025.


\bibitem{MultimodalFinAgent} W. Zhang, L. Zhao, H. Xia, \emph{et al.}, ``A multimodal foundation agent for financial trading: Tool-augmented, diversified, and generalist,'' in \textit{Proceedings of the 30th ACM SIGKDD Conference on Knowledge Discovery and Data Mining}, pp. 4314--4325, 2024.

\bibitem{FromLLMtoAIAgents} M. A. Ferrag, N. Tihanyi, and M. Debbah, ``From LLM reasoning to autonomous AI agents: A comprehensive review,'' \textit{arXiv preprint} arXiv:2504.19678, 2025.


\bibitem{FromNewsToForecast} X. Wang, M. Feng, J. Qiu, \emph{et al.}, ``From news to forecast: Integrating event analysis in LLM-based time series forecasting with reflection,'' \textit{Advances in Neural Information Processing Systems}, vol. 37, pp. 58118--58153, 2024.



\bibitem{RandomWalk} B. G. Malkiel, \textit{A Random Walk Down Wall Street: The Time-Tested Strategy for Successful Investing}. New York: W. W. Norton \& Company, 2019.

\bibitem{NonRandomWalk} A. W. Lo and A. C. MacKinlay, ``A non-random walk down Wall Street,'' in \textit{A Non-Random Walk Down Wall Street}. Princeton, NJ: Princeton University Press, 2011.

\bibitem{BigData22} Y. Soun, J. Yoo, M. Cho, \emph{et al.}, ``Accurate stock movement prediction with self-supervised learning from sparse noisy tweets,'' in \textit{Proceedings of the 2022 IEEE International Conference on Big Data (Big Data)}, pp. 1691--1700, 2022.



\bibitem{ACL18} Y. Xu and S. B. Cohen, ``Stock movement prediction from tweets and historical prices,'' in \textit{Proceedings of the 56th Annual Meeting of the Association for Computational Linguistics (Volume 1: Long Papers)}, pp. 1970--1979, 2018.

\bibitem{CIKM18} H. Wu, W. Zhang, W. Shen, \emph{et al.}, ``Hybrid deep sequential modeling for social text-driven stock prediction,'' in \textit{Proceedings of the 27th ACM International Conference on Information and Knowledge Management}, pp. 1627--1630, 2018.

\bibitem{TAbook} J. J. Murphy, \textit{Technical Analysis of the Financial Markets: A Comprehensive Guide to Trading Methods and Applications}. New York: Penguin, 1999.

\bibitem{IsAllInfo} B. Wang, G. Johnson, M. Hybinette, \emph{et al.}, ``Is all the information in the price? LLM embeddings versus the EMH in stock clustering,'' \textit{arXiv preprint} arXiv:2509.01590, 2025.

\bibitem{Du2024LLMFinSent}
K.~Du, F.~Xing, R.~Mao, and E.~Cambria,
``An evaluation of reasoning capabilities of large language models in financial sentiment analysis,''
in \textit{2024 IEEE Conference on Artificial Intelligence (CAI)}, pp.~189--194, 2024.


\bibitem{TradingAgents} Y. Xiao, E. Sun, D. Luo, \emph{et al.}, ``TradingAgents: Multi-agents LLM financial trading framework,'' \textit{arXiv preprint} arXiv:2412.20138, 2024.

\bibitem{WhatDoesChatGPT} S. Chen, T. C. Green, H. Gulen, \emph{et al.}, ``What does ChatGPT make of historical stock returns? Extrapolation and miscalibration in LLM stock return forecasts,'' \textit{arXiv preprint} arXiv:2409.11540, 2024.

\bibitem{SorosReflexivity}
G.~Soros, ``Fallibility, reflexivity, and the human uncertainty principle,'' \textit{Journal of Economic Methodology}, vol.~20, no.~4, pp.~309--329, 2013.


\end{thebibliography}
\end{document}